\documentclass[amsmath,amssymb,onecolumn, a4paper,14pt]{revtex4}
\usepackage{amsmath}
\usepackage{amssymb}
\usepackage{amsmath,amsthm,amssymb,amscd}
\usepackage{latexsym}
\usepackage{indentfirst}
\usepackage{subfigure}
\usepackage{graphicx}
\usepackage{extarrows}
\usepackage{bm}
\usepackage{multirow}
\usepackage[colorlinks,linkcolor=blue,hyperindex,dvipdfm]{hyperref}

\date{\today}

\begin{document}

\title{Translation with frameshifting of ribosome along mRNA transcript}

\author{Jingwei Li and Yunxin Zhang} \email[Email: ]{xyz@fudan.edu.cn}
\affiliation{Shanghai Key Laboratory for Contemporary Applied Mathematics,
Centre for Computational Systems Biology, School of Mathematical Sciences, Fudan University, Shanghai 200433, China. \\
Shanghai Center for Mathematical Sciences,  Shanghai 200433, China.}

\begin{abstract}
Translation is an important process for prokaryotic and eukaryotic cells to produce necessary proteins for cell growth. Numerious experiments have been performed to explore the translational properties. Diverse models have also been developed to determine the biochemical mechanism of translation. However, to simplify the majority of the existing models, the frameshifting of ribosome along the mRNA transcript is neglected, which actually occurs in real cells and has been extensively experimentally studied. The frameshifting of ribosome evidently influences the efficiency and speed of translation, considering that the peptide chains synthesized by shifted ribosomes will not fold into functional proteins and will degrade rapidly. In this study, a theoretical model is presented to describe the translational process based on the model for totally asymmetric simple exclusion process. In this model, the frameshifting of the ribosome along the mRNA transcript and the attachment/detachment of the ribosome to/from the main body of mRNA codons during translation elongation process, are explicitly included.
The results show that, with ribosome frameshifing, the speed of correctly synthesized peptide chains may increase first and then decrease with both the translation initiation rate $\alpha$ and the ribosome detachment rate $\omega_d$. This results indicates that regulating the translation process to reach maximal synthesized speed of proteins is theoretically feasible. Traffic-related problems of ribosome motion along the mRNA transcript are also addressed theoretically. Depending on parameter values, shock wave (or domain wall) may exist for ribosome probabilities along the mRNA.
\end{abstract}


\maketitle

\section{Introduction}
\begin{figure}
  \centering
  \includegraphics[width=15cm]{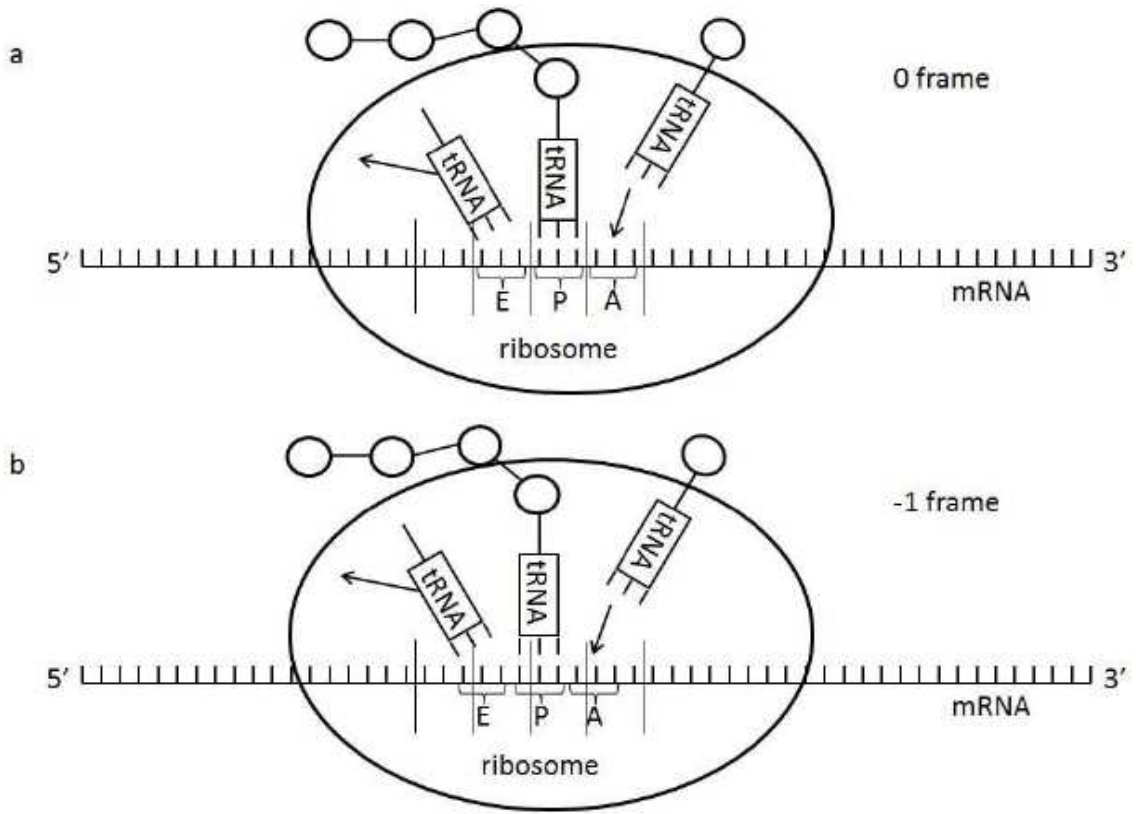}\\
  \caption{{\bf (a)} Normal translation process, in which the ribosome is always bound in the 0 frame of the mRNA. With one forward hopping of ribosome, a new amino acid carried by tRNA is added to the tail of peptide chain. {\bf (b)} Incorrect translation, in which ribosome is bound in the $-1$ frame of mRNA. Compared with the normal ribosome in {\bf (a)}, the ribosome in {\bf (b)} shifts one nucleic acid upstream along the mRNA (see the long vertical lines as reference position.) 
  }\label{frameshift2}
\end{figure}

In living cells, proteins are synthesized by ribosome through the translation process with the information coded in the template messenger RNA (mRNA). The mRNA is composed of a sequence of codons, and each codon consists of three nucleotides. Given the four kinds of nucleotides in mRNAs, a total of $4^3=64$ kinds of codons exits. Except for the stop codon, which is used to finish the translation process, each kind of codon corresponds to a certain kind of amino acid, although each kind of amino acid may correspond to several kinds of codons \cite{Turanov2009}. During translation, an amino acid is transferred by the transfer RNA (tRNA). With the binding of tRNA to mRNA through one of its end, which has complementary anticodon sequence to that of the mRNA, the amino acid carried by the other end of tRNA is then chained together into a polypeptide as the mRNA passes through. A schematic description of this process is found in Fig. \ref{frameshift2}.

Translation is followed by molecular machine ribosome, which includes a three step process, that is, initiation, elongation, and termination \cite{Racle2013}. The 30S subunit of the ribosome contains three tRNA binding sites, which are denoted by A, P, and E (Fig. \ref{frameshift2}). In the initial step, the first tRNA and the start codon of mRNA will form a codon-anticodon duplex in the P site of the 30S subunit, with the Shine-Dalgarno (SD) sequence on the 5' end of the mRNA binding with the 3' end of the 16S rRNA (an anti-SD sequence in the 30S subunit of a ribosome) \cite{Shine1974, Czernilofskya1975, Yusupova2001,Korostelev2007, Duval2013}. Meanwhile, the 50S subunit of ribosome will attach to the mRNA from the other side. During translation, mRNA will be kept in the channel between these two subunits. During elongation, each binding site of the 30S subunit of ribosome will be bound by one mRNA codon, while the tRNA anticodon end will bind to the mRNA codon in the A site. When the ribosome steps forward along the mRNA, the codon from the A site, together with the tRNA attached to it, will move to the P site. Then, the amino acid carried by the tRNA will be added to the tail of the synthesizing peptide chain in the P site. Finally, when the complex of mRNA codon and tRNA moves to the E site, the tRNA will detach from the mRNA codon and the peptide chain (Fig. \ref{frameshift2}). No tRNA that contains an anticodon that is complementary to the stop codon exists. In the termination process, the stop codon will be bound by one release factor, which helps the release of the peptide chain and ribosome from the mRNA, which completes the translation process.

The translation process described above is only applicable for ideal cases, in which the ribosome always moves forward codon by codon, and it always reads three nucleic acids from the same codon in each elongation step. However, in actual cells, the ribosome may shift to one nucleic acid upstream or downstream along the template mRNA in a low frequency \cite{Atkins2010}. This phenomenon is called frameshifting. Two mRNA codons are considered to be in the same frame if and only if the number of nucleic acids between theses codons is divisible by 3. In the following, the mRNA start codon is called a 0 frame codon (or a codon in 0 frame). Thus, all codons that are in the same frame with the start codon are also 0 frame codons (or codons in 0 frame). A codon that is one nucleic acid upstream to a 0 frame codon is called a $-1$ frame codon (or a codon in $-1$ frame). Meanwhile, a codon that is one nucleic acid downstream to a 0 frame codon is called a $+1$ frame codon (or a codon in $+1$ frame). Considering that every codon consists of three nucleic acids, any mRNA codon must belong to one of the three frames, i.e 0, $-1$, and $+1$ frames. If one mRNA codon read by ribosome belongs to the 0 frame (or $\pm$1 frame), then the mRNA is said to be read in the 0 frame (or $\pm$1 frame). Therefore, each mRNA has three reading frames. With frameshifting of ribosome, the reading frames of mRNAs are distinct. To obatin intuitive impressions of frameshifting, one schematic diagram of $-1$ frameshifting is depicted in Fig. \ref{frameshift2}.

Translation is generally initialized from the start codon, so, in the initiation of translation, mRNA is usually in the 0 frame. During the elongation period, frameshifting may occur at some codons, so the mRNA will change into the $-1$ frame or $+1$ frame. Regardless of the reading frame, translation will continue until the ribosome meets the stop codon. However, peptide chains produced by ribosome in $\pm1$ reading frames are different from those produced by correct translations in the 0 reading frame, and these chains are usually nonfunctional, and will degrade rapidly.

Experimental findings have shown that the frequency of frameshifting is only $10^{-3}$ to $10^{-4}$ per codon. However, the frequency of $-1$ frameshifting increases for a slippery sequence with the motif X XXY YYZ \cite{Chen2014}. The 3' secondary structure and the 5' internal SD sequence in the mRNA can further increase the $-1$ frameshift frequency \cite{Larsen1994, Larsen1997}. Numerous studies have been performed to describe the property and mechanism of frameshifting \cite{Tinoco2013,Plant2003,Baranov2004,Horsfield1995,Lopinski2000,Jacks1988,Namy2006,Weiss1989,Leger2007}. A general mechanistic and conformational framework for $-1$ frameshifting is recently presented in \cite{Chen2014} to attempt to elucidate the mechanism of frameshifting. Structural insights into $+1$ frameshifting are discussed in \cite{Maehigashi2014}, and some possible mechanisms of frameshifting can also be found in \cite{Belew2014}.

Biophysically, translation can be roughly regarded as a totally asymmetric simple exclusion process (TASEP) see \cite{Derrida1993,Parmeggiani2003,Zhang20101,Zhang2012}. TASEP is one statistical physics model to describe the unidirectional hopping process of general particles along a one-dimensional lattice. In this model, the particle at site $i$ of the main body of lattice hops forward to site $i+1$ provided that site $i+1$ is not occupied. The particle at the last site of lattice  leaves the lattice into the environment at a given rate constant, and particles in the environment enter the first site of lattice provided that this site is unoccupied. If one mRNA with $N+1$ codons is considered as a one-dimensional lattice of length $N+1$, then the initiation process of translation corresponds to the entrance of the ribosome into the first lattice site. The elongation process of translation corresponds to the forward hopping process of ribosome along the main body of lattice, and finally the termination of translation corresponds to the leaving of ribosome from the last lattice site. Therefore, to a certain extent, TASEP is a reasonable model to describe the translation process in gene expression.

Different types of TASEP models have been conceptualized in various kinds to describe corresponding biophysical or biochemical processes. In these models, particles may be allowed to detach from or attach to any site of lattice, particles may have multiple internal states, there may be different types of particles which hop along the same track, particles may be allowed to hop to the nonadjacent lattice sites, and particles may hop along multiple parallel lattices  \cite{Krug1991,Kolomeisky1998,Schutz2003,Lipowsky2006,Raguin2013,Popkov2013,Bressloff2013,Zhang20113,Zhang20131,Gupta2014}.
Nevertheless, no theoretical model to date has been presented to study the translational process with occasional frameshifting of ribosome along mRNA.  Frameshifting influences on both the speed of correct translation (i.e., the product speed of correct peptide chain) and the accuracy of translation. Therefore, to determine the detailed translational properties, and explore the mechanism to regulate gene expression, a more reasonable model that includes frameshifting should be presented. In this study the translation process, which includes initiation, elongation and termination, will also be regarded as one TASEP of particle hopping along a one-dimensional lattice. Contrary to the usual TASEP, particles (i.e., ribosomes) in the translation elongation period may make one frameshifting stochastically. The corresponding peptide chains produced by shifted ribosomes will be nonfunctional and degrade quickly. Given the difference of ribosomes in correct and incorrect translation states, the proposed model is similar to the TASEP with two particle species. However, the proposed model is different because, during forward hopping, particles may change from one species to the other. Additionally, due to  frameshifting, ribosomes in incorrect translation state do not stop at the supposed end of lattice. This is because that for ribosomes that translocate in the wrong mRNA frames (i.e., $\pm$1 frame), the stop codon cannot be recognized as usual. This study was primarily designed to discuss the translational properties of by the modified TASEP model, which includes the frameshifting of ribosome, especially the dependence on model parameters of correct translation speed and the ratio of correctly produced peptide chains in all peptide chains.

This study is organized as follows. The theoretical model will be presented in the next section, and then the numerical results on the translational properties with ribosome frameshifting will be provided in Section III. Finally, concluding remarks will be presented in the last section.

\section{Modified TASEP with ribosome frameshifting}
\begin{figure}
  \centering
  \includegraphics[width=15cm]{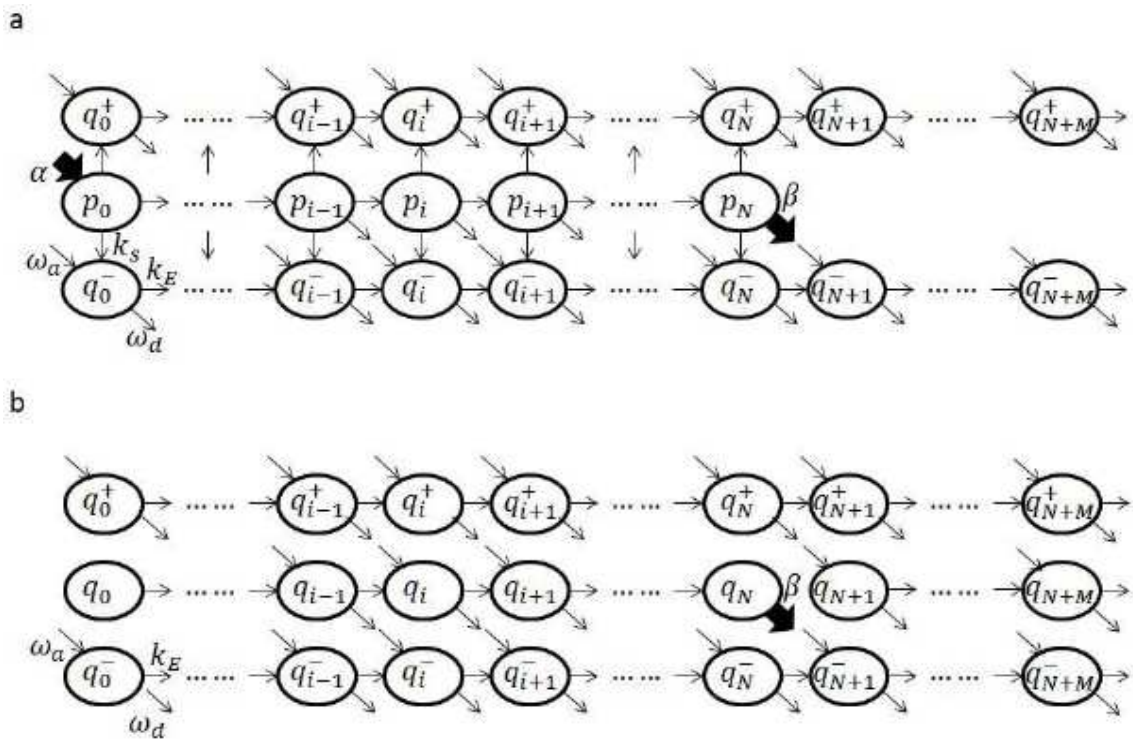}\\
  \caption{Modified TASEP to describe translation with ribosome frameshifting. The initiation of translation begins with the entrance of ribosome to the start codon, which corresponds to the left boundary of the lattice with site index 0. In any mRNA codon $i$ (lattice site $i$), ribosomes may have four possible states. $p_i$ in {\bf (a)} is the probability of the existence of a ribosome at codon $i$ with a correctly synthesized peptide chain, while $q^{\pm}_i$ are the probabilities of the presence of a ribosome at codon $i$ of mRNA in $\pm1$ frame. Given that ribosomes may bind to any mRNA codon, the peptide chain produced by a ribosome in 0 frame may not be correct. For example, the ribosome at codon $i$ may begin its translation from codon $1\le j\le i$. Thus, the peptide chain synthesized by it is also incorrect. The probability of the existence of such a ribosome at codon $i$ is denoted by $q_i$, see {\bf (b)}. Ribosomes in the 0 frame will leave mRNA from the stop codon [the codon $N$ at the right boundary of the middle lattice line in {\bf (a)}]. Generally, ribosomes in $\pm1$ frame will not stop their translation processes until they reach the 3' end of mRNA. $M$ denotes the number of mRNA codons between the stop codon and 3' end of mRNA. In the figures, $\alpha$ is the initiation rate of translation, $\beta$ is the termination rate of translation, and $\omega_a$ and $\omega_d$ are the attachment and detachment rates of a ribosome to and from mRNA, respectively. The elongation rate is denoted by $k_E$, which is assumed to be the same for ribosomes in any mRNA frame and at any codon. Finally, the rate of frameshifting, to either $+1$ frame or $-1$ frame, is assumed to be the same and denoted by $k_s$.
  }\label{model}
\end{figure}

In the usual TASEP without ribosome frameshifting and attachment/detachment to/from the main body of mRNA, the model parameters that influence translation speed are as follows: {\bf (i)} entrance rate of particles from the environment to the first lattice site, i.e., the rate of ribosome binding to the mRNA start codon. This value depends on the concentration of free ribosomes, and the sequence of ribosome binding site (RBS); {\bf (ii)} length of lattice (mRNA), which is usually denoted by $N$; {\bf (iii)} leaving rate $\beta$ of particles (ribosomes) from the last lattice site $N$ (the stop codon) to environment; and {\bf (iv)} forward hopping rate $k_{E}$ of particles in the main body of lattice, i.e., the elongation rate of translation. For simplicity, the particles are usually assumed to be able to hop forward along the track, and their forward hopping rates at any lattice site $i$ are assumed to be the same. In this study, the usual TASEP had been modified to include frameshifting of ribosome as shown in the schematic depiction in Fig. \ref{model}.

Given that the frequency of frameshifting is usually low, this study assumes that throughout the translocation along mRNA, each ribosome can only frameshift at most once. This assumption implies that only ribosomes in the 0 frame can shift one nucleotide forward to 1 frame or backward to $-1$ frame. General models without this restriction can be presented similarly, but those models will complicate the following analysis. Actually, with more possible frameshitings or additional ribosome frameshiftings in the $\pm1$ frame, the results obtained in this study will not change essentially. In this study, the codon with one nucleotide upstream or downstream of codon $i$ is denoted by $i-$ or $i+$. The rates of frameshifting from codon $i$ to codon $i-$ and codon $i+$ are denoted by $k_{-}$ and $k_{+}$, respectively.

Considering that ribosomes only have one chance of framshifting, the ribosomes in $\pm1$ frame will not shift back into the 0 frame. Thus, no stop codon will generally exist in the $\pm1$ frame, and ribosomes in the $\pm1$ frame will come across the stop codon and continue moving forward to the 3' end of mRNA. This study denotes the number of codons between the usual stop codon and the 3' end of mRNA by $M$, and they are numbered by $(N+1)\pm, \cdots, (N+M)\pm$. Thus, the length of track for ribosomes translocating in $\pm1$ frame is $N+M$, which is $M$ codons longer than that for ribosomes translocating in the 0 frame (Fig. \ref{model}). The hopping rate of particles (elongation rate of translation) from codon $i\pm$ to codon $(i+1)\pm$ are denoted by $k^{\pm}_{E}$ and the probabilities of finding one ribosome at codon $i\pm$ are denoted by $q^{\pm}_i$.

In addition to start and stop codons, this study also allows the attachment/detachment of ribosomes to/from any mRNA codon with rates denoted by $\omega_{a}\slash \omega_{d}$. Attachment and detachment of ribosomes are also allowed at codon $i\pm$ for any $0\le i\le N+M$ with rates denoted by $\omega^{\pm}_{a}$ and $\omega^{\pm}_{d}$ respectively. Therefore, ribosomes at codon $i$ for $1\le i\le N$ can be divided into two classes, namely, ribosomes with correctly and incorrectly synthesized peptide chains. If a ribosome reaches codon $i$ from the start codon $0$ without frameshifting, then the synthesized peptide chain will be correct. On the contrary, if a ribosome binds to mRNA at codon $j$ for $1\le j\le i$, then the synthesized peptide chain will be incorrect. For $0\le i\le N$, the probability of finding a ribosome at codon $i$ with one correct or incorrect peptide chain is denoted by $p_i$ or $q_i$. This study assumes that only ribosomes in the 0 frame with a correct peptide chain can frameshift, and the probability $q_0$ is always equals to zero. Considering that frameshifting of ribosomes with incorrect peptide chain will not change the correct translation speed essentially. Without loss of generality, except for the normal initiation and termination, this study assumes the absence of ribosome detachment/attachment at the start/stop codon.

For simplicity, this study assumes that $k_E=k^{\pm}_E, \omega_a=\omega^{\pm}_a, \omega_d=\omega^{\pm}_d$ and $k_+=k_-=:k_s$. This condition indicates that for a ribosome regardless of frames, either in the 0 or in $\pm1$ frame, its forward translocation rate and detachment/attachment rates are the same. Meanwhile, the rates of upstream and downstream frameshifting are also equal.

At any time, only one of the three codons $i$, $i+$ and $i-$ can be occupied by ribosome. Thus, if codon $i$ is occupied by one ribosome, then the codons $i\pm$ will be empty. This phenomenon indicates that the ribosome at codon $i$ always has the possibility to frameshift and then translocate to either codon $i-$ or codon $i+$. For convenience, $W_i$ denotes the probability that position $i$ is unoccupied. One can show that (see Fig. \ref{model}),
\begin{equation}\label{eq1}
W_i=\left\{
            \begin{array}{ll}
                1-q^+_i-q_i-q^-_i-p_i, \quad & 0\le i\le N, \\
                1-q^+_i-q_i-q^-_i, & N+1\le i\le N+M. \\
            \end{array}
      \right.
\end{equation}
The probabilities $p_i$ of finding ribosome with {\it correctly} synthesized peptide chain at codon $i$ are governed by the following equations:
\begin{equation}\label{eq2}
\begin{array}{lll}
{dp_0}/{dt}&=-2k_sp_0+\alpha W_0-k_Ep_0W_1,  & \\
{dp_i}/{dt}&=-2k_sp_i+k_Ep_{i-1}W_{i}-k_Ep_iW_{i+1}-\omega_dp_i, \quad & 1\le i\le N-1, \\
{dp_N}/{dt}&=-2k_sp_N+k_Ep_{N-1}W_{N}-\beta p_N. &
\end{array}
\end{equation}
Similarly, the equations for probabilities $q_i^{\pm}$ of finding ribosome at codon $i\pm$ are as follows:
\begin{equation}\label{eq3}
\begin{array}{lll}
{dq^{\pm}_0}/{dt}&=k_sp_0-k_Eq^{\pm}_0W_1+\omega_aW_0-\omega_dq^{\pm}_0, &  \\
{dq^{\pm}_i}/{dt}&=k_sp_i+k_Eq^{\pm}_{i-1}W_{i}-k_Eq^{\pm}_iW_{i+1}+\omega_aW_i-\omega_dq^{\pm}_i, \quad & 1\le i\le N, \\
{dq^{\pm}_i}/{dt}&=k_Eq^{\pm}_{i-1}W_{i}-k_Eq^{\pm}_iW_{i+1}+\omega_aW_i-\omega_dq^{\pm}_i, & N+1\le i\le N+M. \\
\end{array}
\end{equation}
Where $W_{N+M+1}=1$. If $M\neq0$ then the probabilities $q_i$ of finding ribosome at codon $i$ but with {\it incorrectly} synthesized peptide chain satisfy
\begin{equation}\label{eq4}
\begin{array}{lll}
{dq_i}/{dt}&=k_Eq_{i-1}W_{i}-k_Eq_iW_{i+1}+\omega_aW_i-\omega_dq_i, &\quad \textrm{for } 1\le i\le N+M \text{ and } i\neq N, N+1, \\
{dq_N}/{dt}&= k_Eq_{N-1}W_{N}-\beta q_N, &  \\
{dq_{N+1}}/{dt}&=-k_Eq_{N+1}W_{N+2}+\omega_aW_{N+1}-\omega_dq_{N+1}, &  \\
\end{array}
\end{equation}
and similar equations can be easily obtained for $M=0$. With the values of probabilities $p_i, q_i$ and $q_i^{\pm}$, the {\it effective} rate of translation initiation can be obtained by $\alpha_{eff}:=\alpha(1-p_0-q_0-q_0^+-q_0^-)$, and the {\it effective} translation rate, i.e. the rate of synthesizing correct peptide chains, can be obtained by $\beta_{eff}:=\beta p_N$. The ratio of $\beta_{eff}$ to $\alpha_{eff}$, $r:=\beta_{eff}/\alpha_{eff}$, is one index to describe the {\it effectiveness} of translation along given mRNA template. Without ribosome frameshifting, attachment and detachment, the steady state value of $r$ is equal to 1. For such simple cases, each ribosome that successfully begins its translation process from the start codon will finally complete its translation at the stop codon with one correctly synthesized peptide chain.

\section{Results}
\begin{figure}
  \centering
  \includegraphics[width=15cm]{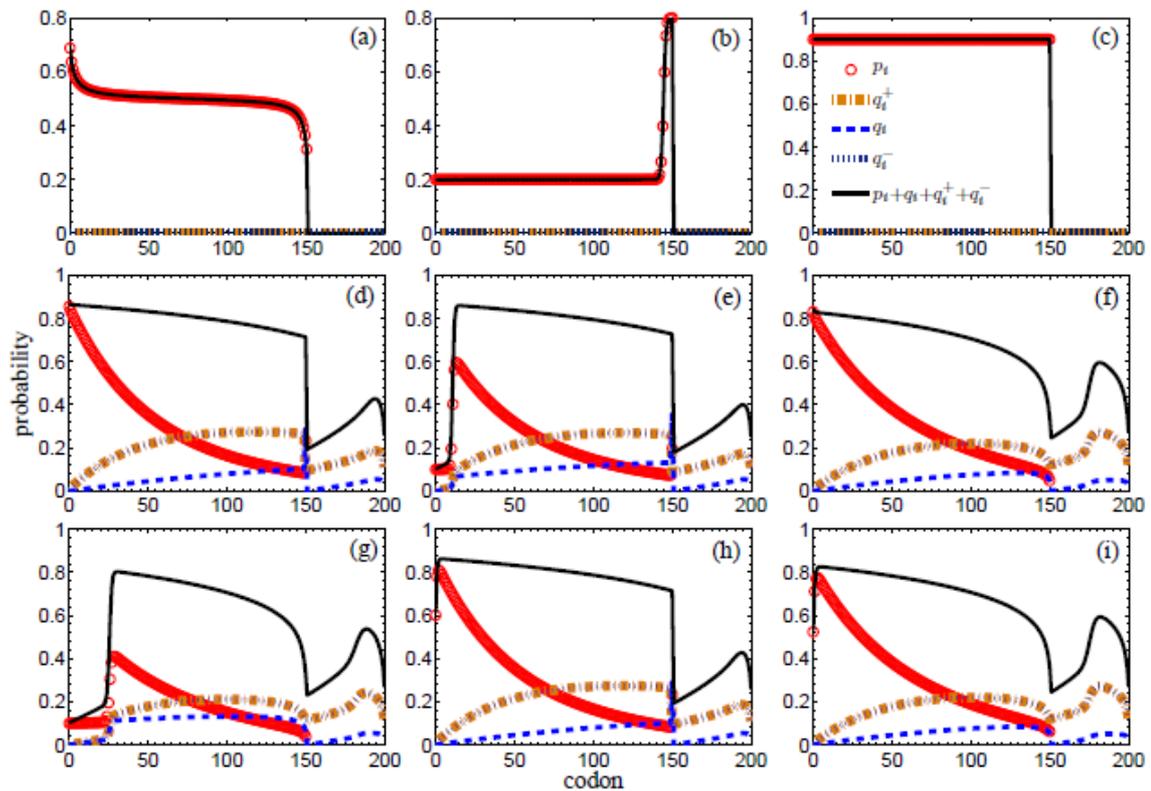}\\
  \caption{Typical examples of probabilities $p_i, q_i, q_i^{\pm}$ and their summation $P_i=p_i+q_i+q_i^{+}+q_i^{+}$ of ribosome along one mRNA with length $N+M=150+50$ (see Fig. \ref{model} for meanings of $N, M$ and the corresponding probabilities.) The parameter values used in these plots are listed in Tab. \ref{probabilitytable}. Roughly speaking, ribosome translocation along the mRNA with frameshifting consists of two TASEPs, which are jointed at lattice site $N$. In {\bf (a,b,c)}, $\omega_a=\omega_d=0$, i.e., except at the start codon and stop codon,  attachment and detachment of ribosome are not allowed. The legend given in {\bf (c)} is right for all these nine figures.
  }\label{probability}
\end{figure}

To illustrate the modified TASEP used in this study for the description of ribosome translocation along mRNA with frameshifting, typical examples of the probabilities $p_i, q_i, q_i^{\pm}$ and their summation $P_i=p_i+q_i+q_i^{+}+q_i^{-}$ are plotted in Fig. \ref{probability}. Where $P_i=1-W_i$ is the probability that site $i$ is occupied by one ribosome. These examples indicate that, from the viewpoint of total probability $P_i$ of ribosome, the modified TASEP can be regarded as a combination of two usual TASEPs for ribosome translocation along mRNA but without frameshifting. Where one involves ribosome translocation from lattice site 0 to lattice site $N$, and the other involves for ribosome translocation from lattice site $N$ to lattice site $N+M$.

For TASEP without ribosome frameshifting, attachment or detachment, i.e., $k_s=\omega_a=\omega_d=0$, the theoretical studies in \cite{Schutz1993,Derrida19931} show a total of three possible phases, namely, low ribosome density phase, high ribosome density phase, and maximal current phase. The plots in Figs. \ref{probability}{\bf (a-c)} show that, such three phases also exist between lattice site $0$ and lattice site $N$ in the modified TASEP. An example for maximal current cases is found in Fig. \ref{probability}{\bf (a)}, in which the total probability $P_i$ is almost equals to 1/2 except for the sharp changes near sites $0$ and $N$. The maximal current phase occurs when both $\alpha$ and $\beta$ are bigger than 1/2. In TASEP, the current is defined by $J=k_EP_i(1-P_i)$. The plots in Figs. \ref{probability}{\bf (b)} and \ref{probability}{\bf (c)} correspond to low and high density phases, respectively, with only sharp change at site $N$ or site $0$. For these special cases, considering the absence of ribosome frameshifting, attachment or detachment, probabilities $q_i, q_i^{\pm}$ are equal to zero, and $p_i=P_i$. The current $j=K_Ep_i(1-p_i)$ of ribosome with correctly synthesized peptide chain is conversed along the mRNA region between the start and stop codons. Thus the {\it effectiveness} $r$ of translation is equal to 1, and actually $\alpha_{eff}$ and $\beta_{eff}$ are both equal to the current $j=K_Ep_i(1-p_i)=k_EP_i(1-P_i)=J$.

For general TASEP, i.e., $\omega_a, \omega_d\ne0$, previous studies have shown that domain wall (or shock wave) and sharp changes at both or only one of the boundaries may exist \cite{Parmeggiani2003, Zhang20101}. The examples plotted in Figs. \ref{probability}{\bf (d-i)} show that, with frameshifting of ribosome,  in the two regions $[0,N]$ and $[N,N+M]$, domain wall and sharp change of probability at one or both of the two boundaries may be found. Meanwhile, the examples plotted in Fig. \ref{probability} show that, similar to the total probability $P_i$, three different phases may also exist for probability $p_i$ (for the cases $\omega_a=\omega_d=0$, see Figs. \ref{probability}{\bf (a-c)}), domain wall and sharp change at boundaries may also exist (for general cases with nonzero attachment and detachment rates, see Figs. \ref{probability}{\bf (d-i)}).
In all these figures, the total probability $P_i$ decreases suddenly at codon $N+1=151$. This phenomenon is due to the existence of the stop codon at site $N=150$. Instead of moving forward to codon $N+1=151$, ribosomes at the stop codon will leave the mRNA template, and consequently the total probability $P_i$ at codon $N+1=151$ will decrease rapidly.

\begin{figure}
  \centering
  \includegraphics[width=15cm]{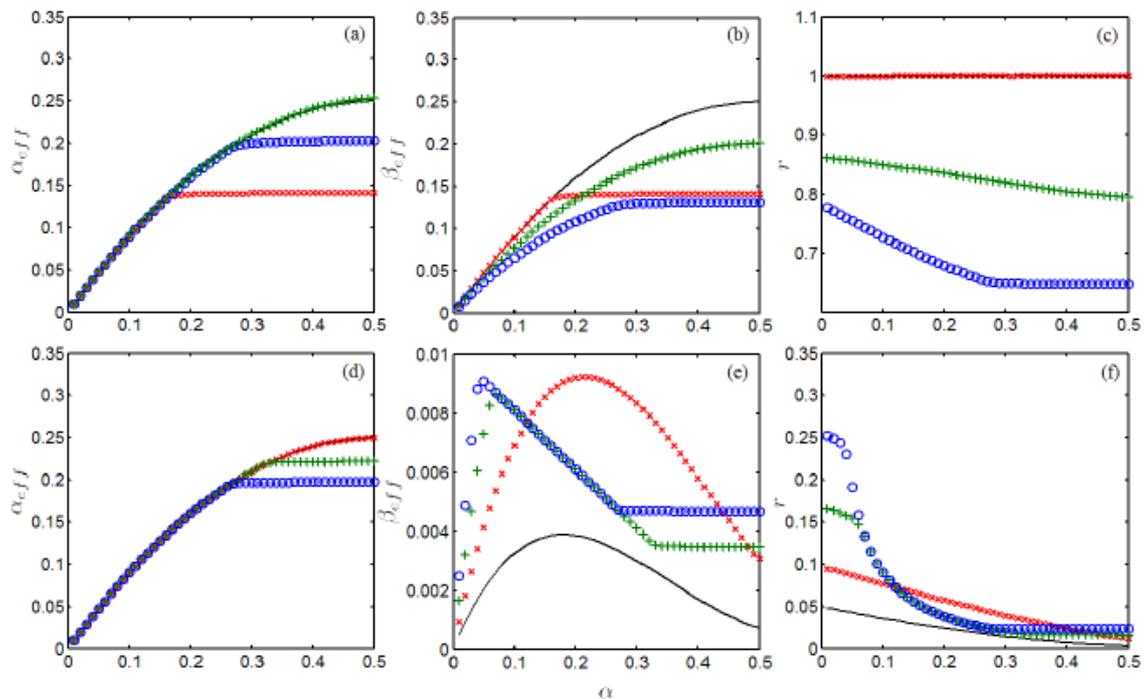}\\
  \caption{The {\it effective} translation initiation rate $\alpha_{eff}:=\alpha(1-p_0-q_0-q_0^+-q_0^-)$ {\bf (a,d)}, {\it effective} (or correct) translation termination rate $\beta_{eff}:=\beta p_N$ {\bf (b,e)}, and their ratio $r:=\beta_{eff}/\alpha_{eff}$  {\bf (c,f)}, as functions of the translation initiation rate $\alpha$. In each figure, four typical examples are plotted, which are drawn by \lq-', \lq$\times$', \lq+' and \lq$\circ$', respectively. I calculations, the initiation rate $\alpha$ changes from 0.01 to 0.5 with an increment of 0.01. The values of other parameters are listed in Table \ref{alphatable}. The main difference between {\bf (a,b,c)} and {\bf (d,e,f)} is that no frameshifting is allowed in {\bf (a,b,c)}, i.e. $k_s=0$, while in {\bf (d,e,f)} ribosomes have nonzero frameshifting rate, i.e., $k_s>0$.
  }\label{alpha}
\end{figure}

\begin{figure}
  \centering
  \includegraphics[width=15cm]{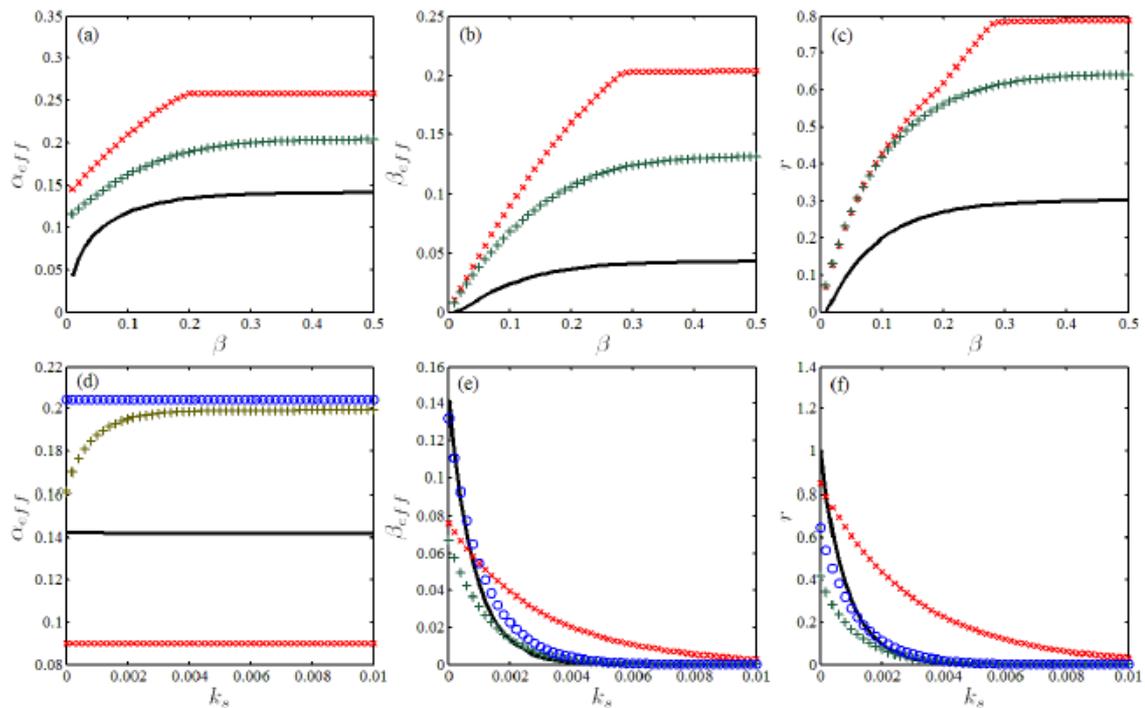}\\
  \caption{The {\it effective} translation initiation rate $\alpha_{eff}$ {\bf (a,d)}, {\it effective} (or correct) translation termination rate $\beta_{eff}$ {\bf (b,e)}, and their ratio $r:=\beta_{eff}/\alpha_{eff}$  {\bf (c,f)}, as functions of the translation termination rate $\beta$ {\bf (a,b,c)}, and the rate of ribosome frameshifting $k_s$ {\bf (d,e,f)}, where $\beta$ changes from 0.01 to 0.5 with an increment of 0.01, and $k_s$ changes from 0 to 0.01 with an increment of 0.002. See Table \ref{alphatable} for other parameter values.
  }\label{betaandks}
\end{figure}

\begin{figure}
  \centering
  \includegraphics[width=15cm]{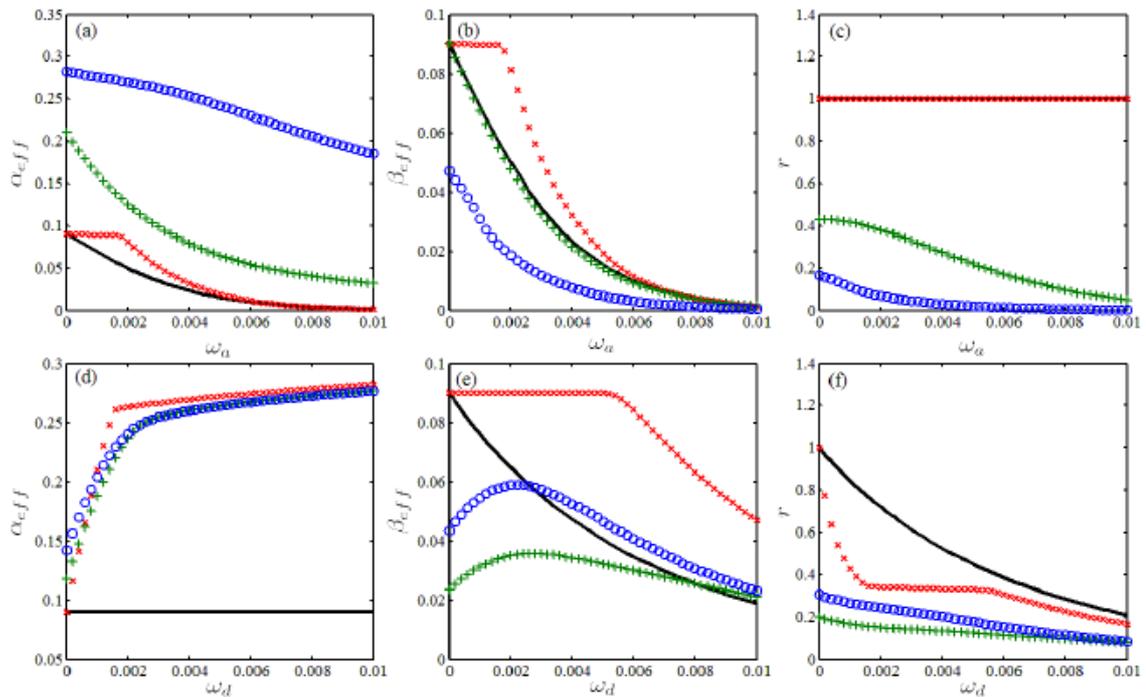}\\
  \caption{The {\it effective} translation initiation rate $\alpha_{eff}$ {\bf (a,d)}, the {\it effective} (or correct) translation termination rate $\beta_{eff}$ {\bf (b,e)}, and their ratio $r:=\beta_{eff}/\alpha_{eff}$  {\bf (c,f)}, as functions of the ribosome attachment rate $\omega_a$ {\bf (a,b,c)}, and the ribosome detachment rate $\omega_d$ {\bf (d,e,f)}, where $\omega_a$ and $\omega_d$ change from 0 to 0.01 with an increment of 0.002. See Table \ref{alphatable} for other parameter values.
  }\label{omegaaandmoegad}
\end{figure}

For translation with ribosome frameshifting, the main properties that should be determined are the influences of frameshifting on the rate of translation initiation $\alpha_{eff}$ and the rate of synthesizing correct peptide chain $\beta_{eff}$, as well as their ratio $r$. Generally, shifted ribosomes, i.e., ribosomes in $\pm1$ frame of mRNA, will not leave the mRNA from the stop codon. The probability of establishing a ribosome at one codon, even if the codon is upstream of the stop codon (i.e., its codon number is less than $N$), may be influenced by the ribosome probabilities between the stop codon and the 3' end of the mRNA. Meanwhile, previous studies about usual TASEP show that ribosome probabilities between sites $N$ and $N+M$ depend on the length $M$ of the untranslated mRNA region \cite{Zhang2012}. Thus, the ribosome probability at any mRNA codon may also depend on parameter $M$, and consequently, any properties of the modified TASEP model for translation with ribosome frameshifting obtained from the probabilities $p_i, q_i$, and $q_i^{\pm}$ may be parameter $M$-dependent. Considering the lack of experimental measured values for parameter $M$, one strategy used in this study is the selection of a sufficiently large value of $M$, such that all the interested properties of translation, especially the {\it effective} initiation rate $\alpha_{eff}$ and termination rate $\beta_{eff}$, reach their steady values. Or in other words, $\alpha_{eff}$ and $\beta_{eff}$ will not change with further increase in parameter $M$, i.e., the length $M$ of the 3' untranslated mRNA region is large enough to have as little influence as possible to the ribosome translocation in the translated region. The numerical calculations show that, for the parameter values used in this study, the selection of one value of $M$ that is larger than 20 is sufficient (Fig. S1 in \cite{supplemental}). In Fig. \ref{probability}, $M=50$ is used, and in Figs. \ref{alpha}-\ref{omegaaandmoegad}, $M=25$ is used.

Without frameshifting, attachment, or detachment, ribosomes initiate their translations from the mRNA start codon and terminate their translations at the stop codon with correctly synthesized peptide chains. On the contrary, with frameshifting, attachment, and detachment, ribosomes may begin/end their translations at any mRNA codon. However, only ribosomes that initiate their translations from the start codon and leave the mRNA from the stop codon synthesize correct peptide chains. Thus, the synthesizing rate of correct peptide chain, which is called the {\it effective} translation rate $\beta_{eff}:=\beta p_N$, will be less than the {\it effective} translation initiation rate $\alpha_{eff}:=\alpha (1-p_0-q_0-q_0^+-q_0^+)$. Their ratio $r:=\beta_{eff}/\alpha_{eff}$ will be one important biophysical index to describe the {\it effectiveness} of gene translation.

In Fig. \ref{alpha}, the {\it effective} translation initiation rate $\alpha_{eff}$, {\it effective} translation speed $\beta_{eff}$, and their ratio $r=\beta_{eff}/\alpha_{eff}$ are plotted with the change in translation initiation rate $\alpha$. The plots in Figs. \ref{alpha}{\bf (a,b)} show that, without frameshifting $k_s=0$, both $\alpha_{eff}$ and $\beta_{eff}$ increase with $\alpha$. By contrast, for nonzero attachment rate $\omega_a$, rates $\alpha_{eff}$ and $\beta_{eff}$ will reach their up limits rapidly. Given that the new ribosome attachment will quicken the saturation of the ribosome on the mRNA. Fig. \ref{alpha}{\bf (c)} shows that, if $\omega_d=0$, then the ratio $r=\beta_{eff}/\alpha_{eff}$ is equal to 1. In such cases, all ribosomes that begin their translations from the start codon will finally end their translations at the stop codon. Generally, $r<1$ holds, and $r$ decreases with initiation rate $\alpha$ and finally tends to one low limit value. For $k_s>0$, i.e., translation with ribosome frameshifting, the plots in Fig. \ref{alpha}{\bf (d)} indicate that the {\it effective} translation initiation rate $\alpha_{eff}$ increases also with $\alpha$, while the plots in Fig. \ref{alpha}{\bf (e)} show that the {\it effective} translation rate $\beta_{eff}$ increases first and then decreases with rate $\alpha$. The reason that $\beta_{eff}$ decreases with large values of $\alpha$ is that for large $\alpha$, the probability of finding a ribosome with incorrect peptide chain will increase, so the translocation speed of ribosome with correct peptide chain will decrease because of the block of ribosome with incorrect peptide chain. Finally, the plots in Fig. \ref{alpha}{\bf (f)} indicate that for $k_s>0$, the ratio $r$ will always be less than 1 and decreases with translation initiation rate $\alpha$. The plots in Fig. \ref{alpha} show that the main differences that resulted from ribosome frameshifting are as follows: the {\it effective} translation rate $\beta_{eff}$ may decrease with translation initiation rate $\alpha$ (see Fig. \ref{alpha}(b,e)), and the {\it effectiveness} $r$ of translation decreases much rapidly with rate $\alpha$ (see Fig. \ref{alpha}(c,f)).

Meanwhile, the plots in Figs. \ref{betaandks}{\bf (a-c)} show that $\alpha_{eff}, \beta_{eff}$, and their ratio $r$ increase with the translation termination rate $\beta$, and tend to corresponding limit constants with large $\beta$. These results imply that, with high leaving rate $\beta$ of ribosome from the stop codon, more correct peptide chains is synthesized, and the {\it effective} initiation rate also increases. Given the increase in leaving rate $\beta$ but fixed rates of frameshifting and detachment, ribosomes with correct peptide chain rapidly translocate along the mRNA template and have less chance to frameshift or detach from the mRNA before they reach the stop codon. On the contrary, the plots in Figs. \ref{betaandks}{\bf (e,f)} indicate that, both $\beta_{eff}$ and ratio $r$ decrease with the rate $k_s$ of frameshifting. Considering large rate $k_s$, ribosomes will have less chance to complete their translation processes correctly, but have more chances to frameshift.
Figs. \ref{omegaaandmoegad}{\bf (b,c)} show that, $\beta_{eff}$ and ratio $r$ also decrease with ribosome attachment rate $\omega_a$. With the increase in attachment rate $\omega_a$, the mRNA template is bound by more ribosomes with incorrect peptide chains, and then the translocation speed of ribosome with correct peptide chain is reduced. Thus, ribosomes with correct peptide chains translocate along the mRNA with lower speed and then have more chances to detach from the mRNA or frameshift before they reach the stop codon. The plots in Fig. \ref{omegaaandmoegad}{\bf (e)} show that $\beta_{eff}$ may not change monotonically with the ribosome detachment rate $\omega_d$. When the detachment rate $\omega_d$ is small, the increase in the rate $\omega_d$ may help decrease the overall ribosome density along the mRNA, and consequently, increase the translocation speed of ribosomes with correct peptide chain. For large values of $\omega_d$, the detachment rate of ribosome with correct peptide chain may increase rapidly, and then the synthesizing rate of correct peptide chain will decrease. This phenomenon implies that, one optimal ribosome detachment rate $\omega_d$ may exist, with which the synthesizing rate of correct peptide chain reaches its maximum.

The plots in Figs. \ref{betaandks}{\bf (d)}, Fig. \ref{omegaaandmoegad}{\bf (a,d)} show that the {\it effective} initiation rate $\alpha_{eff}$ increases with the rate $k_s$ of frameshifting and the rate $\omega_d$ of detachment, but decreases with the attachment rate $\omega_a$. With large detachment/attachment rate, the start mRNA codon has more chances to be unoccupied/occupied, therefore $\alpha_{eff}$ is enlarged/reduced correspondingly. Fig. \ref{betaandks}{\bf (d)} also shows that with large values of frameshifting rate $k_s$, the {\it effective} initiation rate $\alpha_{eff}$ tends to approach one limit constant. With large frameshifting rate $k_s$, the entrance rate of ribosome to mRNA is only determined by the detachment rate $\omega_d$, the attachment rate $\omega_a$, and the termination rate $\beta$. The overall density of the ribosome along the mRNA, with either correct peptide chain or incorrect peptide chain, is completely determined by $\omega_{a}, \omega_d, \alpha$, and $\beta$, see  \cite{Derrida1993,Parmeggiani2003,Zhang20101}. The plots in Fig. \ref{omegaaandmoegad} show that, although the translation {\it effectiveness} $r$ always decreases with attachment rate $\omega_a$ and detachment $\omega_d$, for some special cases the {\it effective} (correct) translation rate $\beta_{eff}$ may increase by increasing the ribosome detachment rate $\omega_d$, see Fig. \ref{omegaaandmoegad}{\bf (e)}. Therefore, detachment maybe one possible mechanism used by cells to increase the synthesizing rate of needed peptide chains.

\section{Conclusions}
Recent experiments found that, during translation, ribosome may frameshift along the mRNA, and the reasons of frameshifting have been experimentally studied thoroughly \cite{Weiss1989,Larsen1994,Horsfield1995,Larsen1997,Jacks1988,Plant2003,Baranov2004,Namy2006,Leger2007,Atkins2010,Tinoco2013,Gupta2013,Chen2014,Belew2014}.
However, so far, no theoretical model has been designed to describe the gene translation process with ribosome frameshifting, especially the influence of frameshifting on translation, such as the {\it effective} translation rate and the efficiency (or {\it effectiveness}) of translation. In this study, a modified TASEP model is presented to describe the translation process with possible ribosome frameshifting. Similar to the usual translation process, i.e., translation without ribosome frameshifting, the probability density of ribosome along the mRNA template may have shock wave (or domain wall) and boundary layers. The {\it effective} translation rate, i.e., the rate of synthesizing correct peptide chain, increases with the translation termination rate $\beta$, and decreases with the rate of frameshifting $k_s$, but may not change monotonically with the translation initiation rate $\alpha$. Meanwhile, the translation {\it effectiveness}, which is defined as the ratio of {\it effective} synthesizing rate of correct peptide chain to the {\it effective} initiation rate of translation, increases with the termination rate, but decreases with initiation rate and the frameshifting rate of ribosome. At the same time, the influences of attachment/detachment of ribosomes to/from the main body of mRNA were also discussed. The results will be beneficial for the understandings of actual translation processes in cells, and the model presented may also be useful for the prediction of translation rate with reasonable chosen parameters.

\begin{acknowledgments}
This study was supported by the Natural Science Foundation of China (Grant No. 11271083), and the National Basic Research Program of China (National \lq\lq973" program, project No. 2011CBA00804).
\end{acknowledgments}

\newpage


\begin{thebibliography}{10}

\bibitem{Turanov2009}
Anton~A. Turanov, Alexey~V. Lobanov, Dmitri~E. Fomenko, Hilary~G. Morrison,
  Mitchell~L. Sogin, Lawrence~A. Klobutcher, Dolph~L. Hatfield, and Vadim~N.
  Gladyshev.
\newblock Genetic code supports targeted insertion of two amino acids by one
  codon.
\newblock {\em Science}, 323(5911):259--261, January 2009.

\bibitem{Racle2013}
Julien Racle, Flora Picard, Laurence Girbal, Muriel Cocaign-Bousquet, and
  Vassily Hatzimanikatis.
\newblock A genome-scale integration and analysis of lactococcus lactis
  translation data.
\newblock {\em PLOS Computational Biology}, 9(10), October 2013.

\bibitem{Shine1974}
J.~Shine and L.~Dalgarno.
\newblock The 3\rq-terminal sequence of {E}scherichia coli 16s ribosomal {RNA}:
  Complementarity to nonsense triplets and ribosome binding sites.
\newblock {\em Proceedings of the National Academy of Sciences of the United
  States of America}, 71(4):1342--1346, April 1974.

\bibitem{Czernilofskya1975}
A.~P. Czernilofskya, C.~G. Kurlanda, and G.~St\"{o}fflerb.
\newblock 30s ribosomal proteins associated with the 3\rq-terminus of 16s
  {RNA}.
\newblock {\em FEBS Letters}, 58(1):281--284, October 1975.

\bibitem{Yusupova2001}
Gulnara~Zh. Yusupova, Marat~M. Yusupov, J.~H.~D. Cate, and Harry~F. Noller.
\newblock The path of messenger {RNA} through the ribosome.
\newblock {\em Cell}, 106(2):233--241, July 2001.

\bibitem{Korostelev2007}
Andrei Korostelev and Harry~F. Noller.
\newblock The ribosome in focus: new structures bring new insights.
\newblock {\em Trends in Biochemical Sciences}, 32(9):434--441, September 2007.

\bibitem{Duval2013}
M\'{e}lodie Duval, Alexey Korepanov, Olivier Fuchsbauer, Pierre Fechter, Andrea
  Haller, Attilio Fabbretti, Laurence Choulier, Ronald Micura, Bruno~P.
  Klaholz, Pascale Romby, Mathias Springer, and Stefano Marzi.
\newblock {E}scherichia coli ribosomal protein {S}1 unfolds structured m{RNA}s
  onto the ribosome for active translation initiation.
\newblock {\em PLOS Biology}, 11(12), December 2013.

\bibitem{Atkins2010}
John~F. Atkins and Raymond~F. Gesteland, editors.
\newblock {\em Recoding: Expansion of Decoding Rules Enriches Gene Expression},
  volume~24 of {\em Nucleic Acids and Molecular Biology}.
\newblock Heidelberg: Springer, New York, 2010.

\bibitem{Chen2014}
Jin Chen, Alexey Petrov, Magnus Johansson, Albert Tsai, Se\'{a}n~E.
  {O\rq}Leary, and Joseph~D. Puglisi.
\newblock Dynamic pathways of $-1$ translational frameshifting.
\newblock {\em Nature}, page to appear, 2014.

\bibitem{Larsen1994}
Bente Larsen, Norma~M. Wills, Raymond~F. Gesteland, and John~F. Atkins.
\newblock r{RNA}-m{RNA} base pairing stimulates a programmed $-1$ ribosomal
  frameshift.
\newblock {\em Journal of Bacteriology}, 176(22):6842--6851, November 1994.

\bibitem{Larsen1997}
Bente Larsen, Raymond~F. Gesteland, and John~F. Atkins.
\newblock Structural probing and mutagenic analysis of the stem-loop required
  for {E}scherichia coli dna{X} ribosomal frameshifting: programmed efficiency
  of 50\%.
\newblock {\em Journal of Molecular Biology}, 271(1):47--60, August 1997.

\bibitem{Tinoco2013}
Ignacio~Tinoco Jr., Hee-Kyung Kim, and Shannon Yan.
\newblock Frameshifting dynamics.
\newblock {\em Biopolymers}, 99(12):1147--1166, December 2013.

\bibitem{Plant2003}
Ewan~P. Plant, Kristi L.~Muldoon Jacobs, Jason~W. Harger, Arturas Meskauskas,
  Jonathan~L. Jacobs, Jennifer~L. Baxter, Alexey~N. Petrov, and Jonathan~D.
  Dinman.
\newblock The 9-{{\AA}} solution: how m{RNA} pseudoknots promote efficient
  programmed $-1$ ribosomal frameshifting.
\newblock {\em RNA}, 9(2):168--174, February 2003.

\bibitem{Baranov2004}
Pavel~V. Baranov, Raymond~F. Gesteland, and John~F. Atkins.
\newblock P-site t{RNA} is a crucial initiator of ribosomal frameshifting.
\newblock {\em RNA}, 10(2):221--230, February 2004.

\bibitem{Horsfield1995}
Julie~A. Horsfield, Daniel~N. Wilson, Sally~A. Mannering, Frances~M. Adamski,
  and Warren~P. Tate.
\newblock Prokaryotic ribosomes recode the {HIV}-1 {\sl gag}-{\sl pol}$-1$
  frameshift sequence by an {E}/{P} site post-translocation simultaneous
  slippage mechanism.
\newblock {\em Nucleic Acids Research}, 23(9):1487--1494, 1995.

\bibitem{Lopinski2000}
John~D. Lopinski, Jonathan~D. Dinman, and Jeremy~A. Bruenn.
\newblock Kinetics of ribosomal pausing during programmed $-1$ translational
  frameshifting.
\newblock {\em Molecular and Cellular Biology}, 20(4):1095--1103, February
  2000.

\bibitem{Jacks1988}
Tyler Jacks, Hiten~D. Madhani, Frank~R. Masiarz, and Harold~E. Varmus.
\newblock Signals for ribosomal frameshifting in the rous sarcoma virus {\sl
  gag}-{\sl pol} region.
\newblock {\em Cell}, 55(3):447--458, November 1988.

\bibitem{Namy2006}
Olivier Namy, Stephen~J. Moran, David~I. Stuart, Robert J.~C. Gilbert, and Ian
  Brierley.
\newblock A mechanical explanation of {RNA} pseudoknot functionin programmed
  ribosomal frameshifting.
\newblock {\em Nature}, 441(7090):244--247, May 2006.

\bibitem{Weiss1989}
R.~B. Weiss, D.~M. Dunn, M.~Shuh, J.~F. Atkins, and R.~F. Gesteland.
\newblock {E}. {\sl coli} ribosomes re-phase on retroviral frameshift signals
  at rates ranging from 2 to 50 percent.
\newblock {\em The New Biologist}, 1(2):159--169, December 1989.

\bibitem{Leger2007}
M\'{e}lissa L\'{e}ger, Dominic Dulude, Sergey~V. Steinberg, and L\'{e}a
  Brakier-Gingras.
\newblock The three transfer {RNA}s occupying the {A}, {P} and {E} sites on the
  ribosome are involved in viral programmed $-1$ ribosomal frameshift.
\newblock {\em Nucleic Acids Research}, 35(16):5581--5592, 2007.

\bibitem{Maehigashi2014}
Tatsuya Maehigashi, Jack~A. Dunkle, Stacey~J. Miles, and Christine~M. Dunham.
\newblock Structural insights into $+1$ frameshifting promoted by expanded or
  modification-deficient anticodon stem loops.
\newblock {\em Proceedings of the National Academy of the Sciences of the
  United States of America}, August 2014.

\bibitem{Belew2014}
Ashton~Trey Belew, Arturas Meskauskas, Sharmishtha Musalgaonkar, Vivek~M.
  Advani, Sergey~O. Sulima, Wojciech~K. Kasprzak, Bruce~A. Shapiro, and
  Jonathan~D. Dinman.
\newblock Ribosomal frameshifting in the {CCR}5 m{RNA} is regulated by mi{RNA}s
  and the {NMD} pathway.
\newblock {\em Nature}, 512(7514):265--269, August 2014.

\bibitem{Derrida1993}
B.~Derrida, S.~A. Janowsky, J.~L. Lebowitz, and E.~R. Speer.
\newblock Exact solution of the totally asymmetric simple exclusion process:
  Shock profiles.
\newblock {\em Journal of Statistical Physics}, 73:813--842, 1993.

\bibitem{Parmeggiani2003}
A.~Parmeggiani, T.~Franosch, and E.~Frey.
\newblock Phase coexistence in driven one-dimensional transport.
\newblock {\em Physical Review Letters}, 90:086601, 2003.

\bibitem{Zhang20101}
Yunxin Zhang.
\newblock Domain wall of the totally asymmetric exclusion process without
  particle number conservation.
\newblock {\em Chinese Journal of Physics}, 48:607--618, 2010.

\bibitem{Zhang2012}
Yunxin Zhang.
\newblock Microtubule length dependence of motor traffic in cells.
\newblock {\em Eur. Phys. J. E}, 35:101, 2012.

\bibitem{Krug1991}
Joachim Krug.
\newblock Boundary-induced phase transitions in driven diffusive systems.
\newblock {\em Phys. Rev. Lett.}, 67:1882--1885, 1991.

\bibitem{Kolomeisky1998}
A.~Kolomeisky, G.M. Schutz, E.B. Kolomeisky, and J.P. Straley.
\newblock Phase diagram of one-dimensional driven lattice gases with open
  boundaries.
\newblock {\em J. Phys. A: Math. Gen.}, 31:6911--6919, 1998.

\bibitem{Schutz2003}
Gunter~M Sch\"{u}tz.
\newblock Critical phenomena and universal dynamics in one-dimensional driven
  diffusive systems with two species of particles.
\newblock {\em J. Phys. A: Math. Gen.}, 36:R339--R379, 2003.

\bibitem{Lipowsky2006}
R.~Lipowsky, Y.~Chai, S.~Klumpp, S.~Liepelt, and M.~J.~I. Muller.
\newblock Molecular motor traffic: From biological nanomachines to macroscopic
  transport.
\newblock {\em Physica A}, 372:34--51, 2006.

\bibitem{Raguin2013}
Ad\'{e}la\"{i}de Raguin, Andrea Parmeggiani, , and Norbert Kern.
\newblock Role of network junctions for the totally asymmetric simple exclusion
  process.
\newblock {\em Physical Review E}, 88:042104, 2013.

\bibitem{Popkov2013}
V.~Popkov, J.~Schmidt, and G.~M. Sch\"{u}tz.
\newblock Superdiffusive modes in two-species driven diffusive systems.
\newblock {\em Physical Review Letters}, 112:200602, 2013.

\bibitem{Bressloff2013}
Paul~C. Bressloff and Jay~M. Newby.
\newblock Stochastic models of intracellular transport.
\newblock {\em Reviews Of Modern Physics}, 85:135--196, 2013.

\bibitem{Zhang20113}
Yunxin Zhang.
\newblock Periodic one-dimensional hopping model with transitions between
  nonadjacent states.
\newblock {\em Phys. Rev. E}, 84:031104, 2011.

\bibitem{Zhang20131}
Yunxin Zhang.
\newblock Theoretical analysis of kinesin {KIF1A} transport along microtubule.
\newblock {\em J. Stat. Phys.}, 152:1207¨C1221, 2013.

\bibitem{Gupta2014}
Arvind~Kumar Gupta and Isha Dhiman.
\newblock Asymmetric coupling in two-lane simple exclusion processes with
  langmuir kinetics: Phase diagrams and boundary layers.
\newblock {\em Physical Review Letters}, 89:022131, 2014.

\bibitem{Schutz1993}
G.~Schutz and E.~Domany.
\newblock Phase transitions in an exactly soluble one-dimensional exclusion
  process.
\newblock {\em Journal of Statistical Physics}, 72:277--296, 1993.

\bibitem{Derrida19931}
B.~Derrida, M.R. Evans, V.~Hakim, and V.~Pasquier.
\newblock Exact solution of a 1d asymmetric exclusion model using a matrix
  formulation.
\newblock {\em J. Phys A: Math. Gen.}, 26:1493--1517, 1993.

\bibitem{supplemental}
The supplementary material including one figure is available at \{URL to be provided\}.

\bibitem{Gupta2013}
Pulkit Gupta, Krishna Kannan, Alexander~S. Mankin, and Nora V\'{a}zquez-Laslop.
\newblock Regulation of gene expression by macrolide-induced ribosomal
  frameshifting.
\newblock {\em Molecular cell}, 52(5):629--642, December 2013.

\end{thebibliography}

\clearpage

\begin{table}
\center
\begin{tabular}{|c|ccccc|}
  \hline
  figures  &\  $\alpha$\  &\  $\beta$\  &\  $k_s$\  &\  $\omega_a$\  &\  $\omega_d$\  \\\hline
  Fig. \ref{probability}{\bf (a)}  & 0.8 & 0.8 & 0 & 0 & 0 \\\hline
  Fig. \ref{probability}{\bf (b)}  & 0.2 & 0.2 & 0 & 0 & 0 \\\hline
  Fig. \ref{probability}{\bf (c)}  & 0.9 & 0.1 & 0 & 0 & 0 \\\hline
  Fig. \ref{probability}{\bf (d)}  & 0.9 & 0.1 & 0.001 & 0.001 & 0.001 \\\hline
  Fig. \ref{probability}{\bf (e)}  & 0.1 & 0.1 & 0.001 & 0.001 & 0.001 \\\hline
  Fig. \ref{probability}{\bf (f)}  & 0.9 & 0.9 & 0.001 & 0.001 & 0.001 \\\hline
  Fig. \ref{probability}{\bf (g)}  & 0.1 & 0.9 & 0.001 & 0.001 & 0.001 \\\hline
  Fig. \ref{probability}{\bf (h)}  & 0.3 & 0.1 & 0.001 & 0.001 & 0.001 \\\hline
  Fig. \ref{probability}{\bf (i)}  & 0.3 & 0.7 & 0.001 & 0.001 & 0.001 \\
  \hline
\end{tabular}
\caption{Parameter values used in Fig. \ref{probability}. In all calculations, $N=150, M=50$ and $k_E=1$ are used. }\label{probabilitytable}
\end{table}

\begin{table}
\center
\begin{tabular}{|c|c|ccccc|}
  \hline
  figures & label & $\alpha$  &  $\beta$ & $k_s$ & $\omega_a$ & $\omega_d$ \\\hline
 & -   &   & 1 & 0 & 0 & 0  \\
   Fig. \ref{alpha}{\bf (a,b,c)} & $\times$    &   & 1 & 0 & 0.001 & 0  \\
    & +   &   & 1 & 0 & 0 & 0.001   \\
    & $\circ$   &   & 1 & 0 & 0.001 & 0.001  \\\hline
& -   &   & 1 & 0.01 & 0 & 0  \\
   Fig. \ref{alpha}{\bf (d,e,f)} & $\times$   &   & 0.05 & 0.007 & 0 & 0   \\
    & +  &   & 0.01 & 0.004 & 0 & 0   \\
    & $\circ$   &   & 0.01 & 0.003 & 0 & 0   \\
    \hline
    \hline
  Fig. \ref{betaandks}{\bf (a,b,c)} & -  & 1 &   & 0.001 & 0.001 & 0  \\
    & $\times$  & 1 &   & 0 & 0 & 0.001  \\
    & +  & 1 &   & 0 & 0.001 & 0.001   \\\hline
  Fig. \ref{betaandks}{\bf (d,e,f)} & -  & 1 & 1 &   & 0.001 & 0  \\
    & $\times$  & 0.1 & 1 &   & 0 & 0.001    \\
    & +  & 1 & 0.1 &   & 0.001 & 0.001    \\
    & $\circ$  & 1 & 1 &   & 0.001 & 0.001   \\
  \hline
  \hline
  Fig. \ref{omegaaandmoegad}{\bf (a,b,c)} & -  & 1 & 0.1 & 0 &   & 0  \\
    & $\times$  & 0.1 & 1 & 0 &   & 0   \\
    & +  & 1 & 0.1 & 0 &   & 0.001   \\
    & $\circ$  & 1 & 0.1 & 0 &   & 0.01  \\\hline
  Fig. \ref{omegaaandmoegad}{\bf (d,e,f)} & -  & 0.1 & 0.1 & 0 & 0 &    \\
    & $\times$  & 1 & 0.1 & 0 & 0 &    \\
    & +  & 1 & 0.1 & 0.001 & 0.001 &      \\
    & $\circ$  & 1 & 1 & 0.001 & 0.001 &      \\
  \hline
\end{tabular}
\caption{Parameter values used in Figs. \ref{alpha}, \ref{betaandks}, and \ref{omegaaandmoegad}. Other parameter values used in the calculations are $N=150, M=25$ and $k_E=1$. }\label{alphatable}
\end{table}

\clearpage
\newpage

\end{document}